# SQUID-based Microtesla MRI for In Vivo Relaxometry of the Human Brain

Vadim S. Zotev, Petr L. Volegov, Andrei N. Matlashov, Igor M. Savukov, Tuba Owens, and Michelle A. Espy

*Abstract*—SQUID-based MRI (magnetic resonance imaging) at microtesla fields has developed significantly over the past few years. Here we describe application of this method for magnetic relaxation measurements in the living human brain. We report values of the longitudinal relaxation time T1 for brain tissues, measured in vivo for the first time at microtesla fields. The experiments were performed at 46 µT field using a seven-channel SQUID system designed for microtesla MRI and MEG. Values of T1, measured for different tissues at this field, are found to be close (within 5%) to the corresponding values of the transverse relaxation time T2 at the same field. Implications of this result for imaging contrast in microtesla MRI are discussed.

*Index Terms* — brain, low-field MRI, relaxation, SQUID

## I. Introduction

Magnetic resonance imaging at microtesla fields using SQUIDs [1], also referred to as ultralow-field (ULF) MRI, is a promising imaging method with some unique and attractive capabilities. ULF MRI is more flexible than conventional MRI, because both polarization of a sample and detection of MRI signals are independent of the Larmor frequency $f$ of nuclear spin precession (for protons, $f/B$=42.6 Hz/µT). In this method, the sample is pre-polarized by a relatively strong (up to 0.1 T and higher) magnetic field prior to each imaging step, and spin precession takes place at a microtesla-range measurement field after the pre-polarizing field is removed [1]. ULF MRI signals are measured using SQUID sensors with untuned input circuits that act as frequency-independent flux-to-voltage converters.

One of the advantages of ULF MRI is a unique opportunity to fully exploit the natural tissue contrast. In conventional MRI, magnetic relaxation occurs at a high magnetic field (e.g., 1.5 T, $f$=64 MHz), which cannot be changed. Most biological tissues, however, exhibit strong magnetic relaxation dispersion, i.e. dependence of the longitudinal relaxation time $T_1$ on magnetic field strength, at lower fields, corresponding to frequencies below ~1 MHz. The dependence of the longitudinal relaxation rate $1/T_1$ on frequency $f$ can be accurately fitted using the following empirical expression [2]:

$$\frac{1}{T_1} = \frac{1}{T_{1,w}} + D + \frac{A}{1+(f/f_c)^{\beta'}} \qquad (1)$$

Here, $1/T_{1,w}$ is the relaxation rate of pure water, and $D$, $A$, $f_c$ and $\beta'$ are free fit parameters [2]. The absolute contrast between two tissues with relaxation times $T_{1a}$ and $T_{1b}$ depends on the difference in their relaxation rates: $1/T_{1a}-1/T_{1b}$. For many tissues, this difference approaches maximum at microtesla fields (typically, $f$ <10 kHz), where the longitudinal relaxation rate $1/T_1$ has a characteristic plateau [3]. Therefore, $T_1$-weighted contrast generally improves as the magnetic field is lowered [4]. The "relative" tissue contrast, however, may reach maximum at medium-range fields (see Fig. 5 in [2]). Because longitudinal relaxation in ULF MRI can be set to take place at any intermediate field between the pre-polarizing field and zero [4], it is possible to selectively adjust $T_1$-weighted contrast between given tissues by manipulating the relaxation field strength. There are hopes that the natural contrast enhancement may enable more efficient cancer screening [5].

The discussion of imaging contrast in ULF MRI would be incomplete without considering the transverse relaxation time $T_2$. The basic theory of NMR relaxation provides the following formulas for isotropic spin ½ systems with dipolar interactions in the motional narrowing regime [3],[6]:

$$\begin{cases} \dfrac{1}{T_1} \approx 6(\omega_{DI})^2 \tau_C \left[ \dfrac{0.2}{1+(\omega\tau_C)^2} + \dfrac{0.8}{1+(2\omega\tau_C)^2} \right] \\ \dfrac{1}{T_2} \approx 6(\omega_{DI})^2 \tau_C \left[ 0.3 + \dfrac{0.5}{1+(\omega\tau_C)^2} + \dfrac{0.2}{1+(2\omega\tau_C)^2} \right] \end{cases} \qquad (2)$$

Here, $\tau_C$ is the correlation time, $\omega=2\pi f$ (where $f$ is the Larmor frequency in an external magnetic field), and $\omega_{DI}$ is the characteristic angular frequency of the dipolar interactions [3]. These formulas predict that $T_1 \to T_2$ as $\omega\tau_C \to 0$, and $T_1 \approx T_2$ in the low-frequency plateau region of $1/T_1$. They also show that the field dependence of $T_2$ is not as strong as that of $T_1$. Therefore, one can expect $T_1$-weighted contrast, established at microtesla fields (i.e. the strongest $T_1$ contrast), to be similar to $T_2$ contrast at microtesla-range and, possibly, higher magnetic fields.

Our group has conducted extensive studies in the field of ULF NMR/MRI for several years [7]-[13]. Recently, we performed imaging of the human brain by ULF MRI in combination with MEG [13]. Here we describe application of ULF MRI for in vivo brain relaxometry. We report $T_1$ values for brain tissues at 46 µT field and compare them to $T_2$ values.

Manuscript received 26 August 2008. This work was supported by the U.S. National Institutes of Health under Grant R01-EB006456 and by the U.S. Department of Energy OBER under Grant KP150302, project ERWS115. The reported study was conducted in compliance with the regulations of the Los Alamos National Laboratory Institutional Review Board for research on human subjects, and informed consent was obtained.

The authors are with Los Alamos National Laboratory, Applied Modern Physics Group, MS D454, Los Alamos, NM 87545, USA (corresponding author V.S. Zotev, phone: 505-665-8460; fax: 505-665-4507; e-mail: vzotev@lanl.gov).

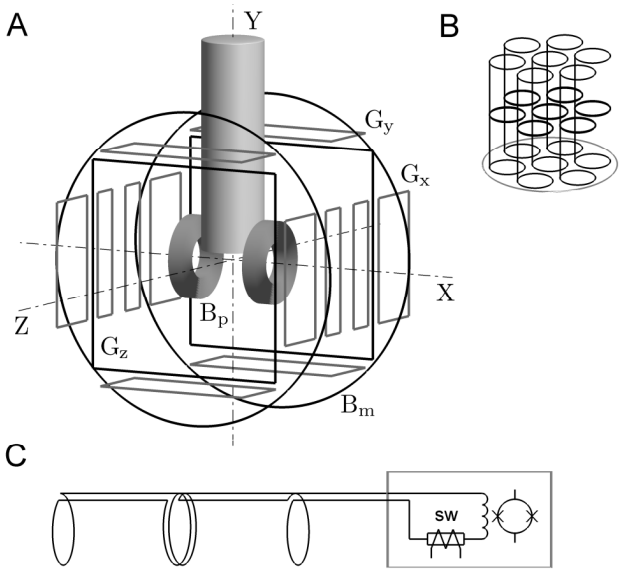

Fig. 1. SQUID-based instrumentation for microtesla MRI and MEG. A) General view of the system. B) Seven second-order gradiometers inside the cryostat. C) Schematic of one SQUID channel.

## II. INSTRUMENTATION

Details of our instrumentation for ULF MRI and MEG have been published before [10]-[13]. The system is depicted schematically in Fig. 1A. It includes a flat-bottom fiberglass liquid helium cryostat with seven second-order SQUID gradiometers, installed inside as shown in Fig. 1B. The gradiometers have 37 mm diameter and 60 mm baseline. They are positioned parallel to one another in a symmetric pattern (Fig. 1B) with 45 mm spacing between centers of the neighboring coils. The field resolution is 1.2 fT/√Hz at 1 kHz for the central channel, and 2.5…2.8 fT/√Hz for the surrounding channels.

A schematic of one SQUID channel is exhibited in Fig. 1C. Special measures have been taken to prevent the SQUIDs from trapping flux during pre-polarization in ULF MRI experiments. Each SQUID sensor is installed inside a sealed lead shield. A cryoswitch, that becomes resistive when a current is applied to its heater, is included in the input circuit between each SQUID and the gradiometer [11]. It is used to effectively disconnect the gradiometer from the SQUID before the pre-polarization, and re-connect it before the measurement.

Our system of magnetic field and gradient coils for ULF MRI, shown in Fig. 1A, includes five sets of coils [13]. A Helmholtz pair generates the measurement field $B_m=46$ μT along Z axis. The pre-polarizing field $B_p=30$ mT is created along X axis by a pair of heavier coils cooled with liquid nitrogen. The two fields are orthogonal in the horizontal (XZ) plane. The other three sets of coils generate gradients for 3D Fourier imaging: $G_z=dB_z/dz$, $G_x=dB_z/dx$, and $G_y=dB_z/dy$. The system is operated inside a magnetically shielded room.

## III. PROCEDURES

The ULF MRI results, presented in this paper, were obtained using two experimental procedures, depicted in Fig. 2. In both cases, the sample was pre-polarized by the $B_p$ field before each imaging step. The field was then ramped down in about 18 ms. Nuclear spin precession was induced by application of the measurement field $B_m$ perpendicular to the original direction of $B_p$ [11] after a few ms delay. The central Larmor frequency at $B_m=46$ μT was about 2 kHz. Phase encoding of the spin precession was performed with two gradients, $G_z$ and $G_y$. Echo signal was then generated by simultaneous reversal of $B_m$ and the frequency encoding gradient $G_x$ [13]. Imaging, therefore, was performed without any rf pulses.

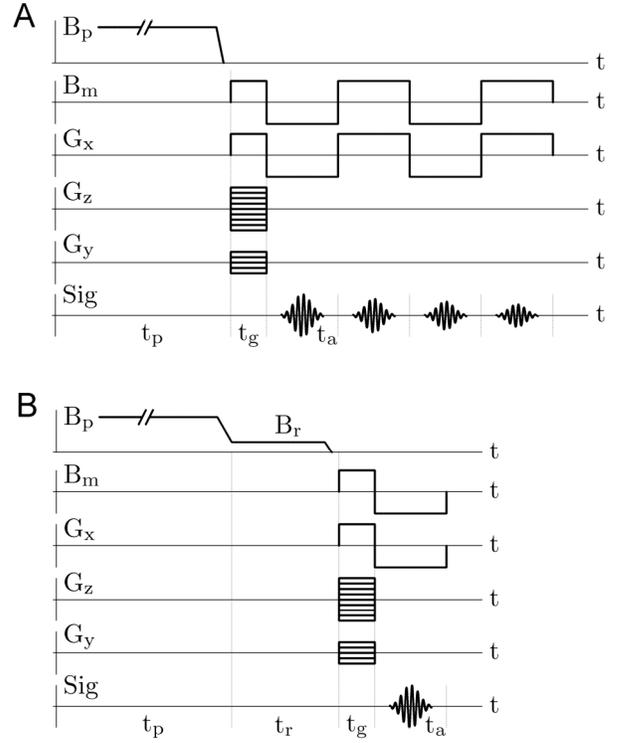

Fig. 2. Experimental procedures for in vivo magnetic relaxometry. A) Protocol for $T_2$ measurements. B) Protocol for $T_1$ measurements (see text).

The multiple-echo procedure in Fig. 2A was used for structural imaging and measurements of the transverse relaxation time $T_2$ [13]. It had the following parameters: $B_p=30$ mT, $B_m=\pm46$ μT, $G_x=\pm140$ μT/m, $|G_z|\leq 140$ μT/m, 61 encoding steps, $|G_y|\leq 70$ μT/m, 11 steps, $t_p=1$ s, $t_g=28$ ms, and $t_a=56$ ms. The imaging resolution in this experiment was 3 mm × 3 mm × 6 mm, with the 6 mm pixel size corresponding to the vertical (Y) dimension. The four echo times, defined as time intervals between the first application of $B_m$ and the top of each echo, were $TE=63, 142, 205$, and 283 ms, respectively.

The experimental procedure in Fig. 2B was used to measure the longitudinal relaxation time $T_1$. For this purpose, it included a variable delay time $t_r$, during which the sample was allowed to undergo longitudinal relaxation in the relaxation field $B_r$. This field was generated by the same $B_p$ coils, but was close in strength to the measurement field $B_m$. It was turned off in about 1 ms. The sequence in Fig. 2B was performed three times with three different delays $t_r$ for any given pair of $G_z$ and $G_y$ values before moving to the next phase encoding step. The following imaging parameters were used: $B_p=30$ mT, $B_r$=46 µT, $B_m$=±46 µT, $G_x$=±140 µT/m, $|G_z| \leq$ 140 µT/m, 31 encoding steps, $|G_y| \leq$ 47 µT/m, 7 steps, $t_p$=1 s, $t_g$=28 ms, and $t_a$=56 ms. The imaging resolution was 3 mm × 3 mm × 9 mm, and the imaging field of view in Z direction was reduced by a factor of 2 from the previous experiment. The three delay times were $t_r$=10, 50, and 90 ms, respectively.

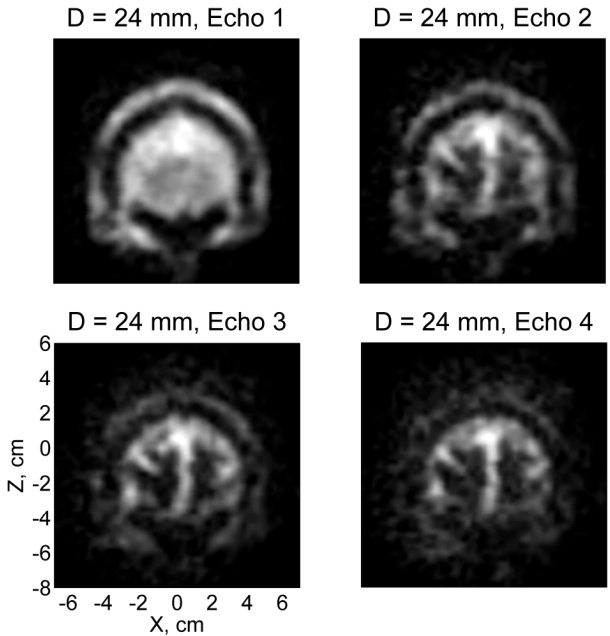

Fig. 3. Images of one 6 mm-thick layer of the human head, corresponding to four consecutive echoes. $D$ is the layer depth under the cryostat bottom.

Six scans of the *k*-space were completed in each of the described experiments, and the resulting images were averaged. The total imaging time was about 90 minutes in each case.

## IV. RESULTS

### A. $T_2$ Measurements

Results of human head imaging by ULF MRI according to the protocol in Fig. 2A are exhibited in Fig. 3. The human subject's head was positioned at the center of the coil system with the forehead against the bottom of the cryostat. Fig. 3 shows images of a 6 mm-thick layer of the head corresponding to four echoes with $TE$=63, 142, 205, and 283 ms, respectively. Only one horizontal image layer out of 11 simultaneously acquired (for each echo) is shown in the figure. Each image is a composite image, obtained as a square root of the sum of squares of images from the seven individual channels.

The multiple-echo technique is commonly used to measure $T_2$ in conventional MRI. To reliably determine $T_2$ from our experimental data, we averaged image intensities for 10-20 pixels, corresponding to a certain tissue type, and fitted the results using a single exponential function, exp(-$TE/T_2$). We obtained the following $T_2$ values at 46 µT field: 106±11 ms for gray matter, 79±11 ms for white matter, 355±15 ms for CSF, and 120±7 ms for scalp. These and other $T_2$ relaxation results for the human head, acquired for the first time at microtesla fields, have been discussed in detail elsewhere [13].

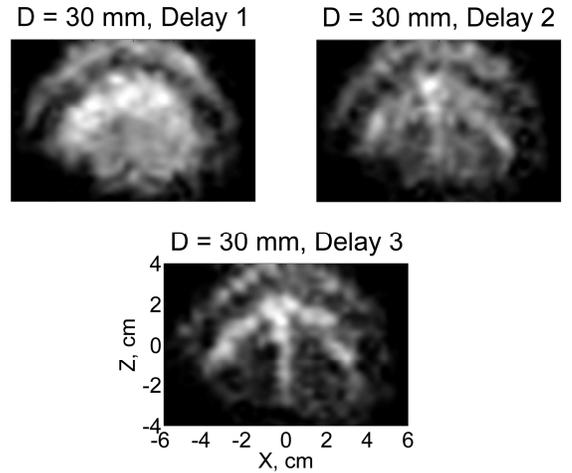

Fig. 4. Images of one 9 mm-thick layer of the human head corresponding to three delay times between the removal of $B_p$ and the application of $B_m$.

### B. $T_1$ Measurements

Measurements of the longitudinal relaxation time $T_1$ are more time consuming than $T_2$ measurements. While the protocol in Fig. 2A allows acquisition of multiple echoes after a single pre-polarization, the procedure in Fig. 2B requires a separate pre-polarization for each delay time $t_r$. Because three different delays were used during the same overall imaging time, the acquisition time for each image was only 1/3 of that in the $T_2$ experiment. This led to a reduction in imaging SNR by a factor of ≈√3 and lower image quality.

Results of the ULF MRI experiment using the protocol in Fig. 2B are exhibited in Fig. 4. This figure shows images of one 9 mm-thick layer of the head (out of 7 simultaneously acquired) for three delay times $t_r$=10, 50, and 90 ms, respectively. The echo time in this experiment was $TE$=63 ms. Because of the limited field of view in Z direction, data from only five channels were included in the composite image.

We used the images for different delays $t_r$ to estimate the longitudinal relaxation time $T_1$ in the human brain for the first time at microtesla fields. The procedure is illustrated in Fig. 5. Regions of gray and white matter were first identified in 3D images of the same human subject's head, acquired by conventional MRI at 1.5 T (see [13]). Then, 10-12 pixels were selected in the corresponding regions of the ULF images, as shown in Fig. 5. CSF and scalp were identified directly. Image intensities for the selected pixels were averaged, and a single exponential function exp(-$t_r/T_1$) was used to fit the data. Noise levels were estimated by averaging intensities of 10 pixels outside the head. We obtained the following $T_1$ values at $B_r$=46 µT relaxation field: 103±5 ms for gray matter, 75±2 ms for white matter, 344±9 ms for CSF, and 124±7 ms for scalp.

The fact that $T_1$ and $T_2$ values, determined from our experimental data, are quite close suggests that, in agreement with (2), the two relaxation times converge in the low-field region for the studied tissues: $T_1 \rightarrow T_2$ as $B_r \rightarrow 0$. One can notice, however, that our $T_1$ values are slightly shorter than the corresponding $T_2$ values in three cases out of four. This discrepancy is not significant, because the relatively large voxel sizes and the limited numbers of data points in our experiments could lead to uncertainties in determination of $T_1$ and $T_2$. Imperfections of the two experimental protocols could also introduce systematic errors in $T_1$ and $T_2$ values. Moreover, the multiexponential nature of transverse relaxation causes $T_2$ decays for gray and white matter to "slow down" at echo times longer than 100 ms (see Fig. 3 in [2]). Note that the error bars for the reported relaxation times specify quality of the exponential fits, and do not characterize reproducibility of the results.

After completing the $T_1$ relaxation analysis of our experimental data, we estimated $T_1$ for gray and white matter using (1). The fit parameters were taken from Table 2 of [2], where they were determined from in vitro NMR relaxation measurements of brain tissues. At $f$=2 kHz, (1) provides $T_1$ estimates of 103 ms and 76 ms for gray and white matter, respectively. The agreement between these numbers and our experimental values of 103±5 ms and 75±2 ms is surprisingly good.

## V. Conclusion

The experimental results, presented in this work, lead to the following conclusions regarding brain imaging by ULF MRI.

First, because $T_1$ and $T_2$ relaxation times for brain tissues become close at microtesla fields, $T_1$-weighted contrast, established at such fields, may be similar to $T_2$ contrast. The exact relation between the two will depend on the proximity of the Larmor frequency in the relaxation field to the low-frequency $1/T_1$ plateau region for each of the tissues involved.

Second, because $T_2$ relaxation time does not exhibit a strong dependence on the magnetic field, $T_1$-weighted contrast, established at microtesla fields, may be similar to $T_2$ contrast, observed at high fields of conventional MRI. For example, the $T_1$ values of 103±5 ms and 75±2 ms for gray and white matter, respectively, reported here at 46 µT field, are not very different from $T_2$ values of 110 ms and 80 ms for the same tissues measured at 3 T field [14].

Third, the main advantage of ULF MRI is not the imaging contrast established at microtesla fields, but the unique opportunity to explore and utilize in vivo $T_1$-weighted contrast as a function of the relaxation field in a wide range of magnetic fields (from zero to 0.1 T and higher). For example, even the type of contrast (positive or negative) between cancerous and healthy tissues may change with the field (see Fig. 7 in [15]).

Fourth, because $T_1$ relaxation times in the human brain at a relatively strong pre-polarizing field are considerably longer than at a microtesla-range measurement field, the duty cycle of ULF MRI becomes less efficient as $B_p$ is increased. For example, one can compare $T_1$≈500 ms for gray matter at 0.1 T [2] and $T_1$≈103 ms for the same tissue at 46 µT.

Finally, further and more accurate studies of in vivo magnetic relaxation by ULF MRI are clearly needed.

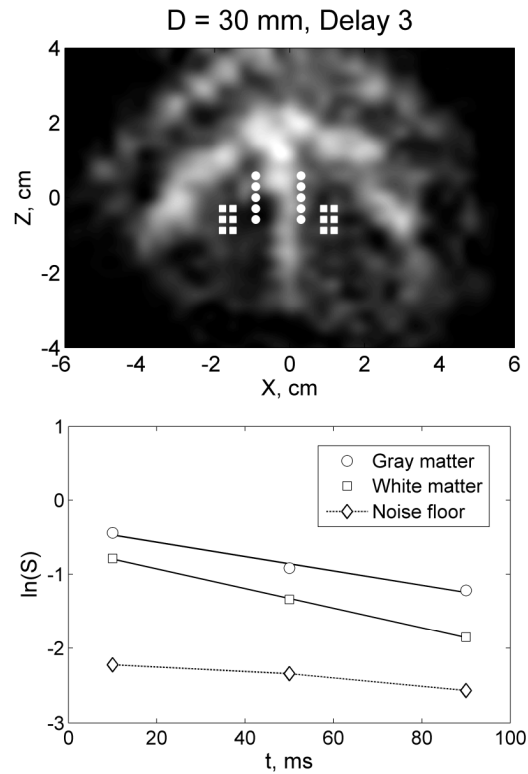

Fig. 5. *Top*: selection of pixels corresponding to gray matter (circles) and white matter (squares). *Bottom*: relaxation data and exponential fits.